\documentclass{article}
\usepackage{caption}
\usepackage{spconf,amsmath,graphicx}
\usepackage{booktabs}
\usepackage{threeparttable}
\usepackage{flushend}
\usepackage{subcaption}
\usepackage{cite}
\usepackage{float}

\title{Data-driven Joint Detection and Localization of Acoustic Reflectors}

\name{H. Nazim Bicer$^{1}$, Cagdas Tuna$^{1}$, Andreas Walther$^{1}$, Emanu\"{e}l A. P. Habets$^{2}$\thanks{$^\dag$A joint institution of the Friedrich-Alexander-Universit\"{a}t Erlangen-N\"{u}rnberg (FAU) and Fraunhofer IIS.
\newline Corresponding author: hasan.nazim.bicer@iis.fraunhofer.de.}}
\address{$^1$Fraunhofer Institute for Integrated Circuits IIS, Erlangen, Germany.
\\ $^2$International Audio Laboratories Erlangen$^\dag$, Germany.}

\begin{document}
\ninept
\maketitle
\begin{abstract}
Room geometry inference algorithms rely on the localization of acoustic reflectors to identify boundary surfaces of an enclosure. Rooms with highly absorptive walls or walls at large distances from the measurement setup pose challenges for such algorithms. 
As it is not always possible to localize all walls, we present a data-driven method to jointly detect and localize acoustic reflectors that correspond to nearby and/or reflective walls. A multi-branch convolutional recurrent neural network is employed for this purpose.
The network's input consists of a time-domain acoustic beamforming map, obtained via Radon transform from multi-channel room impulse responses.
A modified loss function is proposed that forces the network to pay more attention to walls that can be estimated with a small error.
Simulation results show that the proposed method can detect nearby and/or reflective walls and improve the localization performance for the detected walls.
\end{abstract}


\section{Introduction}
\label{sec:intro} 

 
Reflective boundary surfaces have a major influence on the acoustic field inside a room. Knowing the positions of these acoustic reflectors may be advantageous to audio applications including auralization \cite{auralization}, dereverberation \cite{audio-visual-dereverb}, and sound source localization and tracking  \cite{reflection-aware-ssl, reflection-difraction-aware-ssl}. Room geometry inference (RGI) aims at extracting information about the shape of a room by using room impulse responses (RIRs) measured between different arrangements of loudspeakers and microphones \cite{antonacci-2010, microsoft-circ-array, dokmani2013, antonacci-habets-rgi}. 

The arrangements commonly used for RGI can be broadly classified into two categories: i) off-device - the positions of the transducers are unknown \cite{antonacci-habets-rgi, radon_1, tuna, tuna2023datadriven}, and ii) on-device - the relative positions of the transducers are known and are placed relatively close to one another (e.g., in a smart speaker) \cite{microsoft-circ-array, el2019room}. If the relative positions are unknown, they need to be estimated in an initial step \cite{antonacci-habets-rgi, radon_1}. While knowing the relative positions is advantageous, having close space transducers might lead to ill-defined problems and is considered to be more challenging \cite{el2019room}.


Most current RGI methods involve a multi-step process that begins with extracting features from RIRs. These features are then linked to their respective acoustic reflectors, which are essential for estimating the locations and orientations of these reflectors \cite{antonacci-2010, dokmani2013, antonacci-habets-rgi, microsoft-circ-array}. While this detailed control over each processing stage aids in understanding an algorithm's behavior at every point, the separation into distinct steps can lead to sub-optimal results. A more effective approach is to integrate all these steps using deep learning, treating reflector localization as a holistic problem. This approach, focusing on a single, comprehensive solution, typically surpasses multi-step approaches in performance due to its inherent optimization towards a global optimum.

Such a deep learning-based approach was adopted in \cite{tuna2023datadriven} to solve the reflector localization (RL) problem for an off-device setup. Here, a convolutional recurrent neural network (CRNN) was employed to estimate the location and orientation of the sidewalls and the location of the floor and ceiling. Here, the RL problem was formulated as a regression problem. The performance of the method was evaluated with measured data and was found to be on par with the multi-step solution proposed in \cite{tuna}. As the proposed method estimates the localization and orientation of six acoustic reflectors from the RIRs, it implicitly assumes the detectability of all walls in the room. In practice, such an assumption may not always be feasible due to acoustic factors, including propagation-distance attenuation, absorptive surface materials, transducer directivity, low signal-to-noise ratio, and occlusion. These factors could make the localization of walls with high absorption and/or at large distances from the measurement setup very challenging.


To enhance the practicality of the method introduced in \cite{tuna2023datadriven}, we propose an approach for simultaneous detection and localization of acoustic reflectors in a room, utilizing an on-device setup. Building upon the CRNN model from \cite{tuna2023datadriven} and incorporating the insights gained from \cite{adavanne}, our method incorporates an additional branch in the network's architecture to facilitate the detection.
The detection aims primarily to redirect the network's focus from highly absorptive or distant boundary surfaces to highly reflective or nearby walls. Our proposed method, named Attention-based Joint Detection and Localization (A-JDL), utilizes a single-term loss function. In this function, the output from the detection branch is seamlessly integrated into the regression loss of the estimation branch. This integration facilitates a ``self-attention'' mechanism akin to those employed in speech recognition \cite{attention-based}, allowing the network to prioritize walls that can be estimated with a smaller error. This approach eliminates the need for explicit labeling of a wall's detectability during training. To control the detectability, we also propose a regularized loss function. We validated our method using simulated data to i) obtain training data for a wide range of room configurations with varying wall absorption coefficients for the considered problem and ii) systematically analyze its performance.

\section{Problem Formulation}
\label{sec:problem setup}


In this work, we consider rooms with four sidewalls of equal height, perpendicular to both the floor and ceiling, in line with common living quarters \cite{room-dwellings}. This includes rooms with tilted sidewalls, thereby extending our focus beyond the traditional shoe-box model. The recording setup features a circular array with a radius of $r$ and $M$ microphones, and a central loudspeaker that also serves as the origin of the 2D coordinate system. The position of the $m$-th microphone in the array is denoted by $\textbf{p}_m = [r\cos (\frac{2\pi m}{M}) , r\sin (\frac{2\pi m}{M}) ]^\textrm{T}$.

The orientation and location of sidewall $w \in \{1, 2, 3, 4\}$ can be described, respectively, by a unit normal vector $\mathbf{v}_w = [ \cos(\varphi_w), \sin(\varphi_w)]$ with the wall angle $\varphi_w$ w.r.t. the device-centric axes, and the wall distance $d_w$ from the origin. Rather than estimating these quantities separately, we merged them into a single vector to obtain the ground-truth regression labels for a given wall $w$ as  
\begin{align}
    [x_w, \, y_w] &= [d_w  \cos ( \varphi_w ), \,d_w  \sin ( \varphi_w )].  \label{wall_norm_repr}
\end{align}

Our aim is to estimate the wall normal vectors $[x_w, \, y_w]$ for $w \in \{1, 2, 3, 4\}$ given a set of RIRs. In addition, we aim to obtain a detection score per wall, denoted by $0 \leq \hat{\delta}_w \leq 1$.

\section{Proposed Method}
\label{subsec:proposed_method}

This section describes the preprocessing steps, network architecture, and the loss function designed for jointly detecting and localizing acoustic reflectors.

\subsection{Radon transform}
\label{radon-preprocessing}
As a preprocessing step, the Radon transform \cite{radon_1,tuna,tuna2023datadriven} is applied to RIRs to generate an acoustic beamforming map. This map serves as the input for the neural network.

Given a set of time-synchronized microphone signals, let $h_m(n)$ denote the RIR at discrete time sample $n$, recorded between the $m$-th microphone and the loudspeaker. Exploiting the geometry of the fixed source-receiver setup, the initial section of the RIRs up to and including the direct path is discarded from each $h_m(n)$ followed by truncation to obtain RIRs of length $L$.
RIRs are then further processed, as in \cite{tuna2023datadriven, radon_1}, to remove redundant peaks due to the loudspeaker response, which may otherwise introduce spurious peaks on the generated map. Assuming the direct-path amplitude is positive, this is done by zero-clipping the negative parts of the RIRs to obtain $h_m^+(n)$.

The 2D Radon transform \cite{tuna2023datadriven} is then defined as  
\begin{equation}
   R(n, \theta) = \sum_{m = 0}^{M-1}  \, \rho^{(m)}_{n,\theta} h_m^+\left(n - \left \lfloor \Delta^{(m)}_{n, \theta}  \right\rceil \right),
\end{equation}
where $\Delta^{(m)}_{n, \theta}\!=\!\frac{f_s}{c} \left (r_n - \rho^{(m)}_{n, \theta} \right )$ is the corresponding time delay in samples with sampling frequency $f_s$ and speed of sound $c$, ${\rho^{(m)}_{n,\theta} = \| \textbf{q}_{n,\theta} - \textbf{p}_m  \|_2}$ is the propagation path distance between $\textbf{p}_m$ and the point $\textbf{q}_{n,\theta} = [r_n \cos \theta, r_n \sin \theta]^\textrm{T}$ at the distance $r_n$ and the angle $\theta$ from the origin. The rounding operation $\lfloor . \rceil $ is replaced by linear interpolation in practice to reduce errors due to discretization. The computed 2D map has the shape $\Theta \times L$, where $\Theta$ is the number of look directions used for the Radon transform. Finally, the values are normalized to the range of $[-1,1]$. 


\subsection{Network architecture}
\begin{figure}[t!]
\centering
\includegraphics[width=6.5cm]{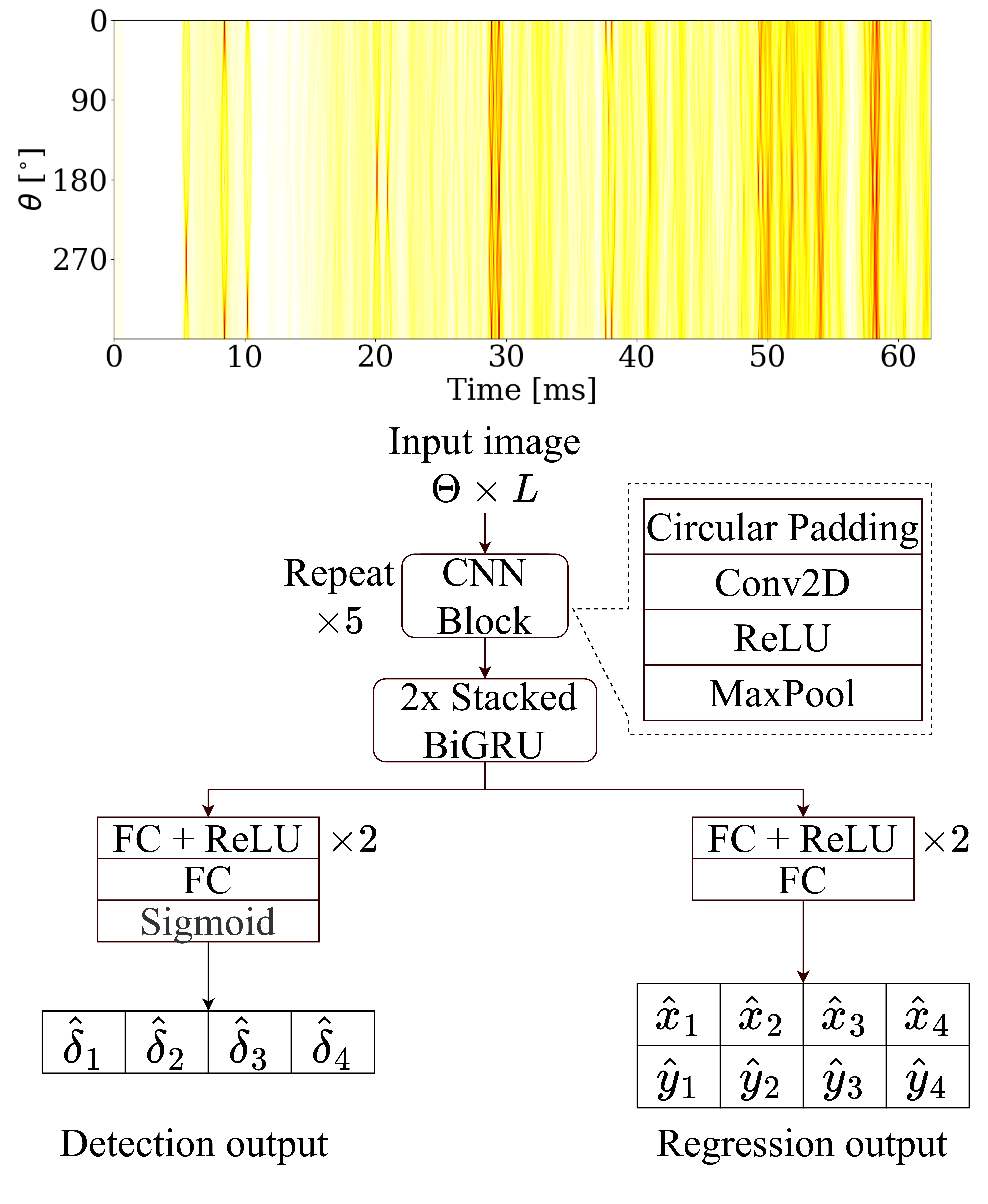}
\caption{CRNN architecture. Detection branch (left) outputs detection scores for each wall, and regression branch (right) estimates corresponding wall normals. The regression branch is equal to the model proposed in \cite{tuna2023datadriven}.} 
\label{fig:CRNN_Model}
\end{figure}

The backbone of the proposed method, depicted in Fig.\,\ref{fig:CRNN_Model}, consists of a CRNN model, followed by fully connected layers as in \cite{tuna2023datadriven}. The single channel input Radon image is initially processed by a sequence of five 2D convolutional blocks. The number of filters are set to $16$, $32$, $64$, $64$, $64$, having the size of $7\!\times\!7$, $5\!\times\!5$, $3\!\times\!3$, $3\!\times\!3$, $3\!\times\!3$, respectively. Max-pooling with a kernel size of $2\!\times\!2$ is applied for dimension reduction across layers.
Prior to each convolutional block, and in contrast to \cite{tuna2023datadriven}, circular padding is applied along the angular dimension, allowing the network to capture the cyclic behavior of the planar array. The output tensor of the fifth convolutional block is fed into the stacked bidirectional Gated Recurrent Units (BiGRUs). The final hidden layers from both the forward and backward passes of the last BiGRU layer are concatenated and given as input to the following detection and estimation branches consisting of fully connected layers. 

\begin{table*}[!t]
\footnotesize
  \centering
  \captionsetup{skip=8pt} 
  \caption{Performance evaluation of the data-driven joint detection and localization (JDL) models. LO(D): LO errors for walls detected by JDL models. LO(U): LO errors for undetected walls by JDL models.} 
  \label{tab:regression}
  \begin{tabular}{l c c c c c c c }
    \toprule
    & & \multicolumn{3}{c}{Distance error [cm]} &\multicolumn{3}{c}{Orientation error [$^\circ$]} \\ \cmidrule(lr){3-5} \cmidrule(lr){6-8}  
    Method & Detection rate [\%] & JDL &  LO(D) & LO(U) & JDL & LO(D) & LO(U) \\ \midrule
     LO (baseline) & -  & - & 16.93 $\pm$ 43.88 & - & - & 3.18 $\pm$ 5.60 & - \\ \cmidrule{1-8}
    A-JDL & 24.58  & \,\,3.27 $\pm$ 18.79 & \,\,4.27 $\pm$ 14.19 & 21.04 $\pm$ 49.17 & 1.80 $\pm$ 3.58 & 2.68 $\pm$ 5.34 & 3.33 $\pm$ 5.66  \\
    RA-JDL ($\lambda=0.01$) & 24.95  & \textbf{0.92} $\pm$ \textbf{3.28} & 2.30 $\pm$ 3.06 & 21.79 $\pm$ 49.68 & \textbf{1.06} $\pm$ \textbf{1.74} & 2.77 $\pm$ 6.85 & 3.31 $\pm$ 5.10 \\
    RA-JDL ($\lambda=0.05$) & 76.27  & \,\,6.68 $\pm$ 25.72 & \,\,7.88 $\pm$ 26.22 & 46.00 $\pm$ 69.23 & 1.94 $\pm$ 3.54 & 2.48 $\pm$ 4.97 & 5.41 $\pm$ 6.79   \\
    RA-JDL ($\lambda=0.10$) & 92.47  & 10.75 $\pm$ 34.41 & 11.85 $\pm$ 34.55 & 79.19 $\pm$ 81.96 & 2.47 $\pm$ 4.23 & 2.79 $\pm$ 5.23 & 7.89 $\pm$ 7.47 \\ \bottomrule
  \end{tabular}
  		    \begin{tablenotes}
      \footnotesize
      \item Detection rate and mean$\pm$std (standard deviation) are computed over 40,000 walls (i.e., 10,000  rooms).
    \end{tablenotes}
\end{table*}

\subsection{Attention-based learning}

The baseline method with no detection capabilities is chosen to be a sub-network of the CRNN shown in Fig.\,\ref{fig:CRNN_Model} that utilizes only the estimation branch, referred to as the Localization-Only (LO) method. The loss function used to train the baseline model is defined as:
\begin{equation}
     L_{\mathrm{LO}} = \frac{1}{4B} \sum_{b=1}^{B} \sum_{w=1}^{4} \ell_w^{(b)},
\end{equation}
where 
\begin{align}
\ell_w^{(b)} &= \sqrt{(x_w^{(b)} - \hat{x}_w^{(b)})^2 + (y_w^{(b)} - \hat{y}_w^{(b)})^2 } \\
& \quad w=1,\dots, 4 \quad b=1,\dots,B, \nonumber
\end{align}
denotes the Euclidean distance between the ground truth and estimated wall normals in (\ref{wall_norm_repr}) for training sample $b$ in the mini-batch and wall $w$.

To enable the neural network to make decisions on wall detection without any supervision, an attention-based loss is proposed. We enable the neural network to self-attend to walls that can be accurately estimated (i.e., with a small error) by incorporating the output of the detection branch in a new loss function, given by
\begin{equation}
L_{\textrm{A-JDL}}=\frac{1}{B}\sum_{b=1}^{B}\frac{\sum_{w=1}^{4} \hat{\delta}_w^{(b)} \ell_w^{(b)}}{\sum_{w=1}^{4} \hat{\delta}_w^{(b)}},
\label{eq:ajdem_loss}
\end{equation}
where ${\hat{\delta}_w^{(b)}}$ is the detection score for training sample $b$ and wall $w$.
The cases when there are no detections, i.e., $\sum_{w=1}^{4} \hat{\delta}_w^{(b)} \approx 0$ are heavily penalized by $L_{\textrm{A-JDL}}$ as it substantially increases. As a consequence, the neural network attempts to detect at least one wall with the smallest Euclidean distance between the ground truth and estimated wall normals, which is most likely to be a nearby and/or highly reflective wall. However, using solely this loss function would result in single-wall detection, following the principle of least effort \cite{shortcut_learning}. 

To overcome this problem, a second loss function, termed \textit{regularized} A-JDL (RA-JDL) loss, with an additional penalty term to allow the detection of up to all four sidewalls, is proposed:
\begin{equation}
L_{\textrm{RA-JDL}}= L_{\textrm{A-JDL}} + \lambda \sqrt{\frac{1}{B} \sum_{b=1}^{B} \left( W_{\max} -  \sum_{w=1}^{4} \hat{\delta}_w^{(b)}   \right)^2},  
\end{equation}
where $\lambda$ is the weight for the penalty term, and $W_{\max}$ is the expected number of detected walls, which is set to $4$ in this study.

\section{Performance Evaluation} 
\label{sec:performance}

This section presents the parameters used during the synthetic RIR generation process, followed by the description of network hyperparameters and metrics used for model evaluation. As a baseline for this study, we use the method proposed in \cite{tuna2023datadriven}, which has been shown to perform on par with the method proposed in \cite{tuna}, which was state-of-the-art at the time of publication.

\subsection{Data, Training and Metrics}

Simulated RIRs were generated using \textit{Pyroomacoustics} \cite{pyroom_acoustics} with the image method \cite{ism}. The highest order of reflections was set to $7$, the sampling frequency ${f_s\!=\!16}$~kHz, and the speed of sound \mbox{${c\!=\!343}$ m/s}. The rooms were simulated using the following steps to generate diverse convex room configurations: A rectangle was first generated with its length and width randomly sampled from the range of $[3\,\mathrm{m},\,8\,\mathrm{m}]$. Each sidewall was then tilted with an angle randomly sampled from $[ -20^\circ, +20^\circ]$. As this process may not guarantee convexity, non-convex 2D floor layouts were discarded. The array had $M\!=\!8$ microphones with $r\!=\!5$~cm. The microphones and the loudspeaker were all assumed to be omnidirectional. The microphone array was randomly positioned within the 2D floor map with its center at a minimum Euclidean distance of $10\,\mathrm{cm}$ from the closest wall, taking into account the microphone array size. The room height $h_\textrm{room}$ was randomly selected from the range of $[2\,\mathrm{m},\,5\,\mathrm{m}]$, and the device height from the floor was varied between $0.5\,\mathrm{m}$ and  $4.5\,\mathrm{m}$. The wall absorption coefficients were randomly sampled from the range of $[0,1)$ for each wall separately.

Overall, $30,000$ rooms were generated for training, and $10,000$ each for validation and test datasets. Pink noise was randomly added to RIRs at signal-to-noise levels ranging from 20 to~50 dB. The Radon maps were computed for $L = 1000$ samples of the  pre-processed RIRs and the angular range of $[0^\circ,360^\circ)$ at $1^\circ$ resolution.  

All neural networks under test were trained for $200$ epochs with a batch size of $B=50$ with AdamW \cite{adamw} optimizer with a weight decay rate of $5\times10^{-5}$. Early stopping was employed with the patience of $20$ epochs. 





The metrics used for performance evaluation are the distance error, $\epsilon_{w,d}$, and the orientation error, $\epsilon_{w,\theta}$, for wall $w$, which are given by \cite{tuna} as
\begin{equation}
\epsilon_{w,d} = | d_w - \hat{d}_w|, \,\, \text{and} \,\, \epsilon_{w,\theta} = \arccos{(< \mathbf{v}_w,\mathbf{\hat{v}}_w >)},  
\end{equation}
where $\hat{d}_w$ and $\mathbf{\hat{v}}_w$ are the estimated distance and the wall unit normal vector, respectively, inferred from the localization branch of the network by using (\ref{wall_norm_repr}):
\begin{equation}
    \mathbf{\hat{v}}_w  = \frac{[\hat{x}_w,\hat{y}_w]}{{|| [\hat{x}_w,\hat{y}_w]||}_2}.
\end{equation}

\subsection{Results}

The performance evaluation of the proposed models is presented in Table \ref{tab:regression}, where LO(D) refers to the LO error for the walls detected by a given JDL model, and contrarily, LO(U) corresponds to the error for the remaining walls undetected by the JDL model.
During inference, a wall detection threshold of $\gamma=0.5$ is employed to determine whether a wall is detected or not by JDL models. The average performances of the JDL models are reported for the detected walls in Table \ref{tab:regression} in the column JDL.
The comparison between LO(D) and LO(U) distance error values indicates that the JDL models were able to distinguish between the walls, but at differing levels, by focusing on the walls that could be localized with smaller errors and neglecting the ones with larger errors.

The attention-based technique improved the localization accuracy in both distance and orientation. A-JDL and RA-JDL ($\lambda=0.01$) reached the lowest wall detection rates, detecting approximately one wall per room and generally the very close and reflective ones with the lowest LO(D) distance errors, in the test dataset. It should also be noted that RA-JDL ($\lambda = 0.01$) outperformed A-JDL considerably (i.e., the mean errors for RA-JDL ($\lambda = 0.01$) were $60\%$ and $61\%$ lower than LO(D) for distance and orientation, respectively), showing the beneficial effect of the regularization term in the RA-JDL loss function. The RA-JDL results demonstrate that the wall detection rates increased as expected with increasing $\lambda$, gradually converging to the case of LO, where all walls were to be localized. Another key observation may also be the considerable increase in the error for the walls undetected by RA-JDL (i.e., LO(U)) with increasing $\lambda$, which may be a result of the regularization term helping the self-attention mechanism in ordering the walls in terms of localization accuracy and leaving the difficult-to-estimate ones out depending on the value of $\lambda$. 


\begin{figure}[!t]
\centering
\begin{subfigure}{0.97\columnwidth}
\centering
\includegraphics[height=4.55cm]{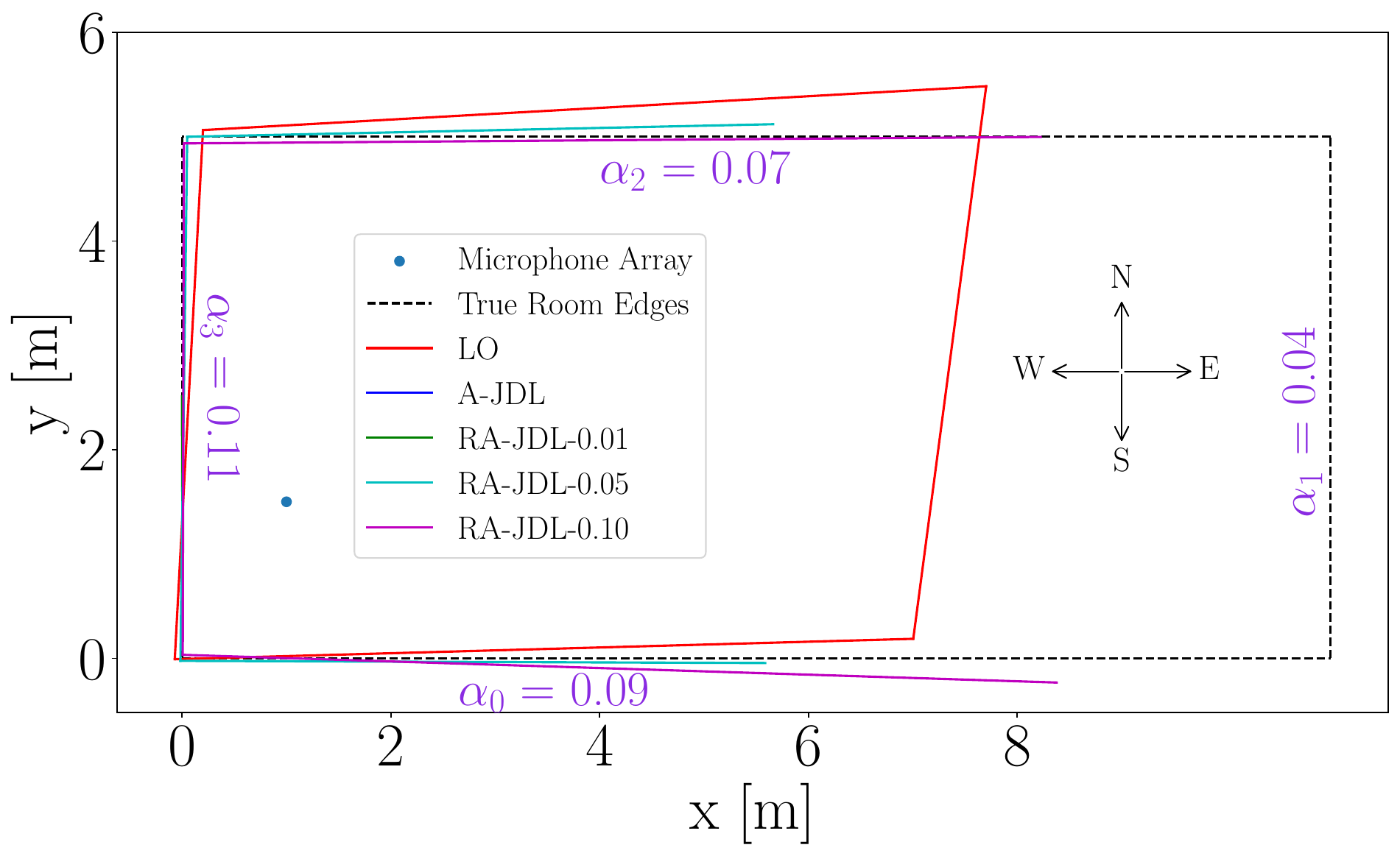}
\caption{Long hallway, out-of-distribution test sample. (legend applies to all subplots).}
\label{fig:long-hallway}
\end{subfigure}
\begin{subfigure}{0.45\columnwidth}
    \centering
    \includegraphics[height=3.cm]{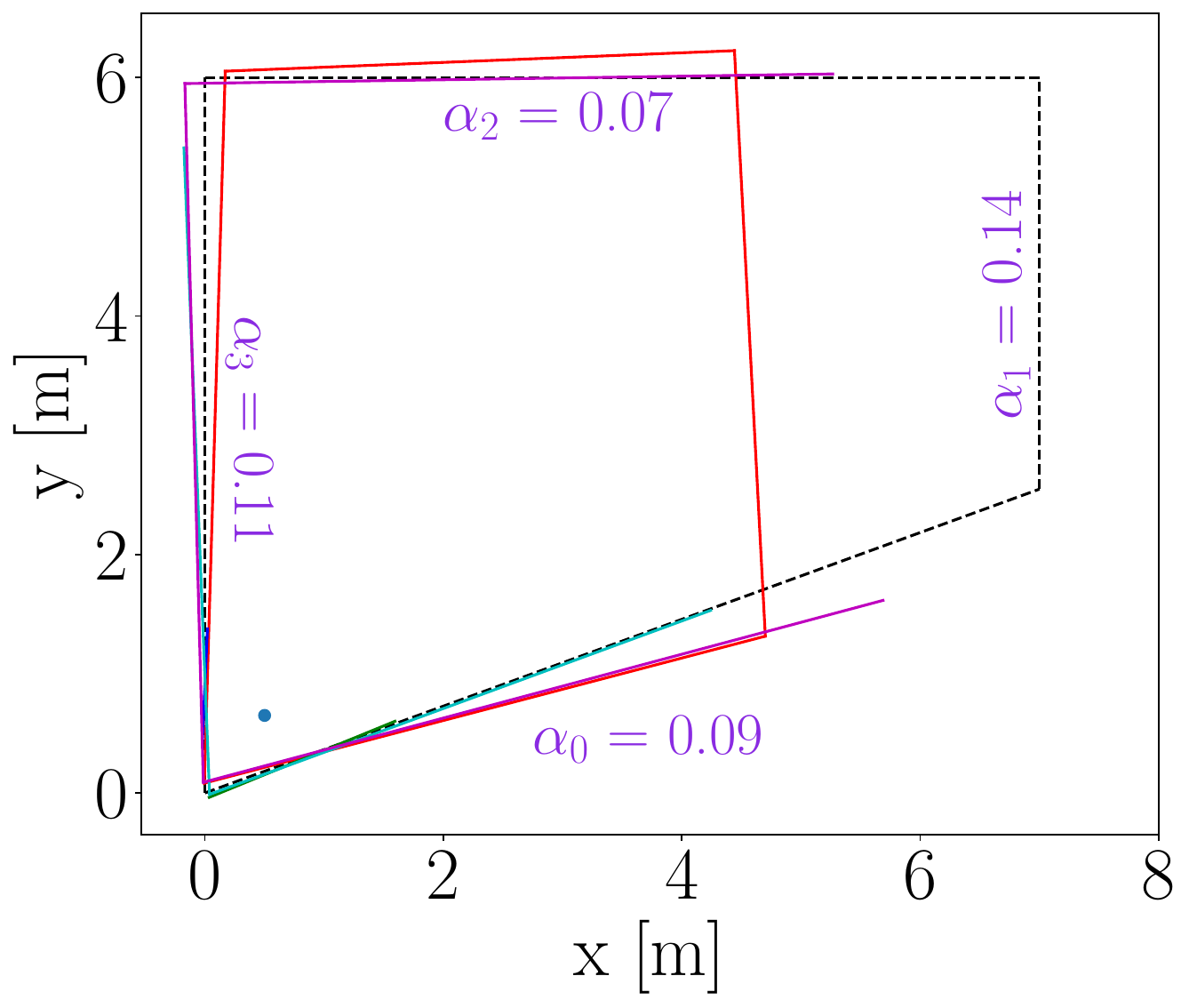}
    \caption{A room with a very tilted wall.}
    \label{fig:unconventional-room}
  \end{subfigure}
  ~
  \begin{subfigure}{0.45\columnwidth}  
    \centering
    \includegraphics[height=3.cm]{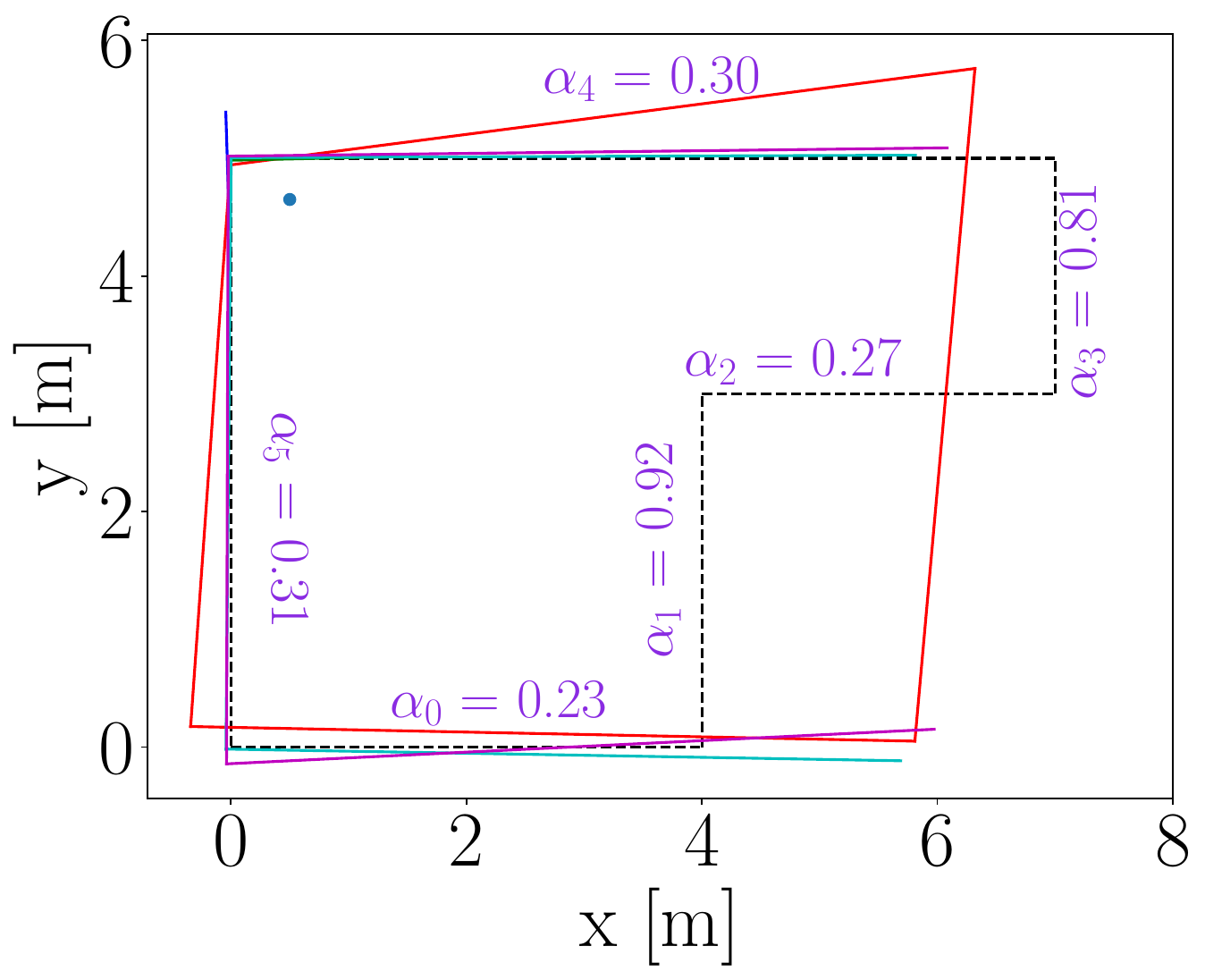}
    \caption{L-shaped room.}
    \label{fig:Lshaped-room}
  \end{subfigure}
  \caption{2D floor map estimation examples.}
  \label{fig:floor_map}
\end{figure}
The inferred 2D floor maps of three examples are shown in Fig.\,\ref{fig:floor_map}. The stark difference between the LO and the JDL models can be observed in Fig.\,\ref{fig:long-hallway}, where an out-of-distribution sample is of interest. The setup was located closer to the west wall of a long hallway, and the far distance between the microphone array and the east wall resulted in large localization errors and poor reconstruction of the complete 2D map for the LO model. On the contrary, all the JDL models yielded partial 2D maps, yet with significantly more accurately localized walls. The second investigated case involved an unconventional room with a very tilted wall, as shown in Fig.\,\ref{fig:unconventional-room}, where the first-order reflection from the east wall was occluded by the south wall, leading to LO generating a phantom east wall due to invisibility.
The third experiment was conducted in an L-shaped room as shown in Fig.\,\ref{fig:Lshaped-room}, where RA-JDL ($\lambda=0.05$) and RA-JDL ($\lambda=0.10$) were able to detect three walls with the lowest errors creating a reasonable partial 2D map.  This behavior may be desirable in large and/or non-convex rooms when compared with the LO model which produces undesired phantom walls.

To analyze the effects of varying wall absorption coefficients for different models, room impulse stacks from a $4 \times 3 \times 3$ shoebox room with reflective walls were simulated. The device is positioned at $[2.25,1.5,0.5]$ and the absorption coefficient of the east wall is varied. In this example, RA-JDL ($\lambda=0.01$) detected the south wall only, possibly due to close distance, while other JDL models showed different behaviors for the detection of the east wall. As expected, models with lower $\lambda$ behaved more conservatively and detected the east wall when it was reflective.

\begin{figure}[t]
\centering
\includegraphics[width=0.8\columnwidth]{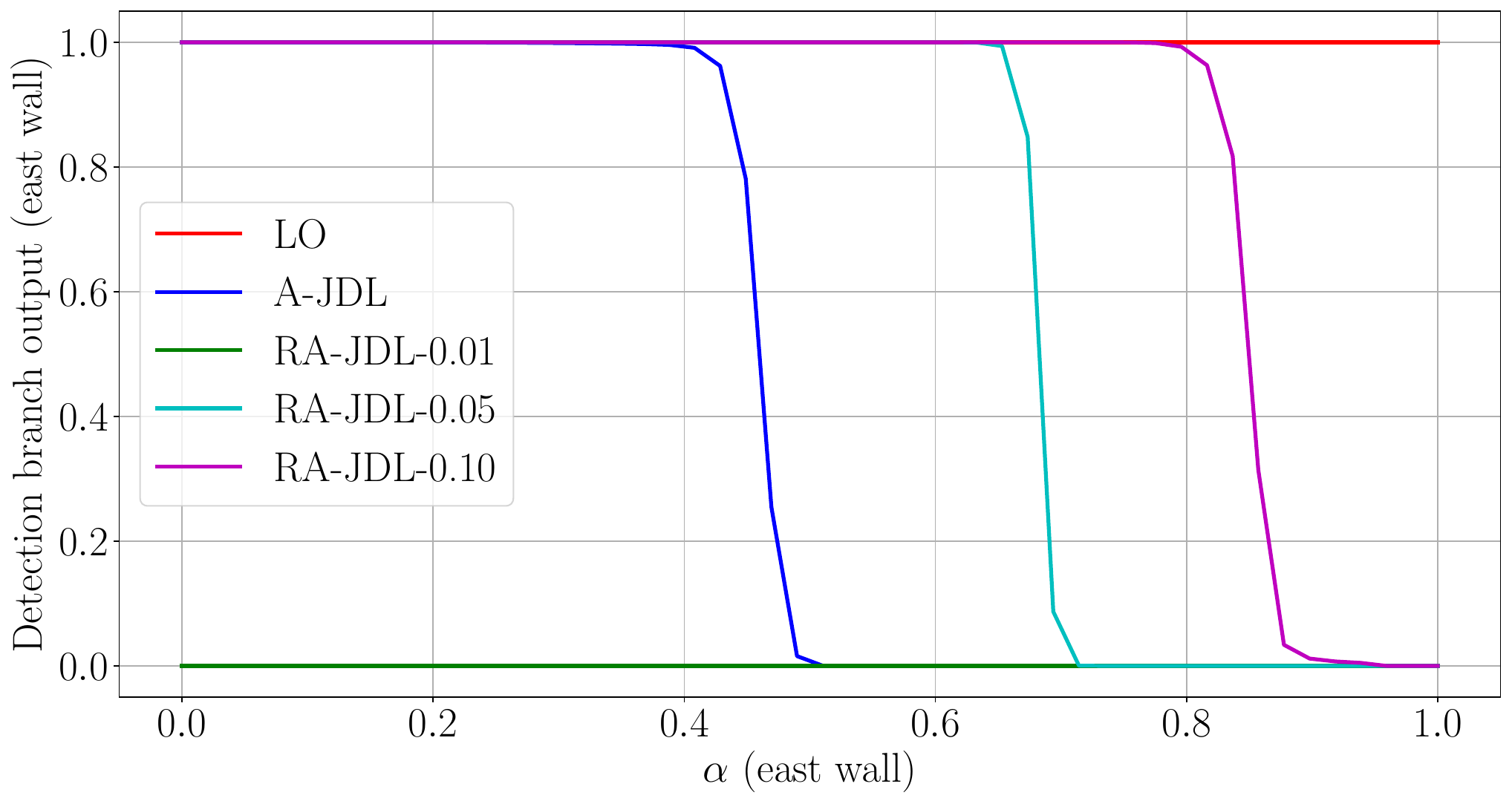}
\caption{Effect of varying absorption coefficient on detection output for different models.}
\label{fig:varying_alpha}
\end{figure}

  ~


\section{Conclusion}
We have introduced a data-driven method with a modified loss function for the joint detection and localization of acoustic reflectors. The proposed method uses a modified loss function to prioritize walls that can be estimated with a small error and does not require supervised learning of the detection scores. It has been shown that the proposed method  generally focuses on walls that can be estimated with high accuracy, while neglecting difficult-to-estimate walls.
The resulting model behavior, where the model effectively focuses on nearby or reflective walls, might be beneficial to analyzing a larger space using multiple devices.
Future work includes training and testing the proposed model with measured data to account for various acoustic phenomena, such as loudspeaker directivity, diffraction, and scattering in rooms with furniture and non-ideal walls.

\section{Acknowledgments}
Parts of this work have been funded by the Free State of Bavaria in the DSAI project.

\bibliographystyle{IEEEbib}
\bibliography{refs}

\begin{thebibliography}{10}

\bibitem{auralization}
M.~Vorl{\"a}nder and J.~Summers,
\newblock ``Auralization: Fundamentals of acoustics, modelling, simulation,
  algorithms, and acoustic virtual reality,''
\newblock {\em J. Acoust. Soc. Amer.}, vol. 123, pp. 4028, July 2008.

\bibitem{audio-visual-dereverb}
C.~Chen, W.~Sun, D.~Harwath, and K.~Grauman,
\newblock ``Learning audio-visual dereverberation,''
\newblock in {\em Proc. {IEEE} Int. Conf. Acoust., Speech Signal Process.}, pp.
  1--5, June 2023.

\bibitem{reflection-aware-ssl}
I.~An, M.~Son, D.~Manocha, and S.~Yoon,
\newblock ``Reflection-aware sound source localization,''
\newblock in {\em IEEE Int. Conf. on Robot. and Automat. (ICRA)}, pp. 66--73,
  2018.

\bibitem{reflection-difraction-aware-ssl}
I.~An, Y.~Kwon, and S.~Yoon,
\newblock ``Diffraction- and reflection-aware multiple sound source
  localization,''
\newblock {\em IEEE Trans. on Robot.}, vol. 38, no. 3, pp. 1925--1944, 2022.

\bibitem{antonacci-2010}
F.~Antonacci, A.~Sarti, and S.~Tubaro,
\newblock ``Geometric reconstruction of the environment from its response to
  multiple acoustic emissions,''
\newblock in {\em Proc. IEEE Int. Conf. Acoust., Speech Signal Process.}, pp.
  2822--2825, 2010.

\bibitem{microsoft-circ-array}
F.~Ribeiro, D.~Florencio, D.~Ba, and C.~Zhang,
\newblock ``Geometrically constrained room modeling with compact microphone
  arrays,''
\newblock {\em IEEE Trans. on Audio, Speech and Lang. Process.}, vol. 20, no.
  5, pp. 1449--1460, 2012.

\bibitem{dokmani2013}
I.~Dokmani{\'c}, R.~Parhizkar, A.~Walther, Y.~M. Lu, and M.~Vetterli,
\newblock ``Acoustic echoes reveal room shape,''
\newblock {\em Proc. of the National Academy of Sciences}, vol. 110, pp. 12186
  -- 12191, 2013.

\bibitem{antonacci-habets-rgi}
F.~Antonacci, J.~Filos, M.~R.~P. Thomas, E.~A.~P. Habets, A.~Sarti, P.~A.
  Naylor, and S.~Tubaro,
\newblock ``Inference of room geometry from acoustic impulse responses,''
\newblock {\em IEEE Trans. on Audio, Speech and Lang. Process.}, vol. 20, no.
  10, pp. 2683--2695, 2012.

\bibitem{radon_1}
Y.~E. Baba, A.~Walther, and E.~A.~P. Habets,
\newblock ``{3D} room geometry inference based on room impulse response
  stacks,''
\newblock {\em IEEE/ACM Trans. on Audio, Speech, and Lang. Process.}, vol. 26,
  no. 5, pp. 857--872, 2018.

\bibitem{tuna}
C.~Tuna, A.~Canclini, F.~Borra, P.~Götz, F.~Antonacci, A.~Walther, A.~Sarti,
  and E.~A.~P. Habets,
\newblock ``{3D} room geometry inference using a linear loudspeaker array and a
  single microphone,''
\newblock {\em IEEE/ACM Trans. on Audio, Speech, and Lang., Process.}, vol. 28,
  pp. 1729--1744, 2020.

\bibitem{tuna2023datadriven}
C.~Tuna, A.~Akat, H.~N. Bicer, A.~Walther, and E.~A.~P. Habets,
\newblock ``Data-driven {3D} room geometry inference with a linear loudspeaker
  array and a single microphone,''
\newblock in {\em Proc. of the Forum Acusticum, Eur. Acoust. Assoc.}, Italy,
  Sep. 11-15 2023,
\newblock Available: {https://arxiv.org/abs/2308.14611}.

\bibitem{el2019room}
Y.~E. Baba, A.~Walther, and E.~A.~P. Habets,
\newblock ``Room geometry inference using sources and receivers on a uniform
  linear array,''
\newblock in {\em Proc. of International Conference on Spatial Audio (ICSA)},
  Germany, pp. 115--121, Sep. 26-28 2019.

\bibitem{adavanne}
S.~Adavanne, A.~Politis, J.~Nikunen, and T.~Virtanen,
\newblock ``Sound event localization and detection of overlapping sources using
  convolutional recurrent neural networks,''
\newblock {\em IEEE Journal of Selected Topics in Signal Process.}, vol. 13,
  no. 1, pp. 34--48, 2018.

\bibitem{attention-based}
J.~Chorowski, D.~Bahdanau, D.~Serdyuk, K.~Cho, and Y.~Bengio,
\newblock ``Attention-based models for speech recognition,''
\newblock in {\em Proc. of the 28th Int. Conf. on Neural Inf. Process. Syst.},
  Cambridge, MA, USA, p. 577–585, 2015.

\bibitem{room-dwellings}
C.~Díaz and A.~Pedrero,
\newblock ``The reverberation time of furnished rooms in dwellings,''
\newblock {\em Applied Acoustics}, vol. 66, no. 8, pp. 945--956, 2005.

\bibitem{shortcut_learning}
R.~Geirhos, J.-H. Jacobsen, C.~Michaelis, R.~Zemel, W.~Brendel, M.~Bethge, and
  F.~A. Wichmann,
\newblock ``Shortcut learning in deep neural networks,''
\newblock {\em Nature Mach. Intell.}, vol. 2, no. 11, pp. 665--673, 2020.

\bibitem{pyroom_acoustics}
R.~Scheibler, E.~Bezzam, and I.~Dokmani{\'c},
\newblock ``Pyroomacoustics: A {Python} package for audio room simulation and
  array processing algorithms,''
\newblock in {\em Proc. IEEE Int. Conf. Acoust., Speech Signal Process.}, pp.
  351--355, Apr. 2018.

\bibitem{ism}
J.~Allen and D.~Berkley,
\newblock ``Image method for efficiently simulating small-room acoustics,''
\newblock {\em J. Acoust. Soc. Amer.}, vol. 65, no. 4, pp. 943--950, Apr. 1979.

\bibitem{adamw}
I.~Loshchilov and F.~Hutter,
\newblock ``Decoupled weight decay regularization,''
\newblock in {\em Int. Conf. Learning Representations}, 2019.

\end{thebibliography}
\end{document}